\documentclass[aps,pra,twocolumn,showpacs,floatfix,superscriptaddress]{revtex4}

\usepackage{amsmath,float,graphicx,subfigure}

\begin{document}

\title{Faraday waves in elongated superfluid fermionic clouds}
\author{P. Capuzzi}
\email{capuzzi@df.uba.ar}
\affiliation{Consejo Nacional de Investigaciones Cient{\'\i}ficas y
T{\'e}cnicas and Departamento de F{\'\i}sica, FCEyN, Universidad de
Buenos Aires, 1428 Buenos Aires, Argentina}
\author{P. Vignolo}
\email{Patrizia.Vignolo@inln.cnrs.fr}
\affiliation{Institut Non Lin\'eaire de Nice,
Universit\'e de Nice-Sophia Antipolis, CNRS,
1361 route des Lucioles,
06560 Valbonne, France}
\begin{abstract}
We use hydrodynamic equations to study the formation of Faraday waves in a 
superfluid Fermi gas at zero temperature confined in a strongly elongated 
cigar-shaped trap. First, we treat the role of the radial density profile
in the limit of an infinite cylindrical geometry and analytically
evaluate the wavelength of the Faraday pattern. The effect of the
axial confinement is fully taken into account in the numerical
solution of hydrodynamic equations and shows that the infinite
cylinder geometry provides a very good description of the phenomena. 
\end{abstract}

\pacs{67.90.+z,03.75.Ss,05.45.-a,47.54.-r}
\maketitle

\section{\label{sec:intro}Introduction}
The term Faraday waves refers to regular surface fringes which are
excited by a vertical oscillatory motion in a nonlinear liquid
\cite{Faraday1831}.  The time-dependent spatially uniform driving
leads {\it via} the nonlinear interaction between the wave excitations
of the system to an instability towards the formation of spatial
structure.  Since the Faraday discovery, the phenomena of pattern
formation and, more generally, of parametric amplification have been
studied in diversified contexts, including convective fluids, nematic
liquid crystals, nonlinear optics and biology \cite{Cross1993}.
Recently, Faraday waves have been experimentally created in a trapped
atomic Bose-Einstein condensate (BEC) by a periodic modulation of the
transverse confinement \cite{Engels2007}.  Nonlinearity in ultracold
gases can be driven either by varying the scattering length, e.g.,
{\it via} Feshbach resonances, as proposed in this context by
Staliunas {\it et al.} \cite{Staliunas2002}, or by varying the trap
parameters as experimentally realized by Engels \textit{et al.}
\cite{Engels2007}.  Several phenomena connected to parametric
amplification of excitations has been theoretically investigated in
atomic BECs \cite{Ripoll1999,Tozzo2005,Modugno2006}
%, as for instance the excitation of axial phonons as a result of the
%periodic modulation of the depth of an optical potential has been
%addressed in \cite{Tozzo2005} while the amplification of
%counterrotating Bogoliubov excitations  in a ring geometry in
%\cite{Modugno2006}.
and very recently, the properties of Faraday waves in confined BECs
has been also explored \cite{Kagan2007,Nicolin2007}. In all these
cases, the periodic modulation of the nonlinearity leads to a
parametric excitation of sound waves in the transverse direction
respect to the modulation.

In this paper, we study the Faraday pattern formation in a superfluid
Fermi gas in the BCS-BEC crossover at zero temperature.  The physics
of Faraday waves in a superfluid Fermi gas is richer than in an atomic
Bose-Einstein condensate because the nonlinearity may also drive the
microscopic features of the superfluid as described by its equation of
state. This leads to strong variations in the pattern along the crossover. 

We adopt the experimental conditions of Engels \textit{et
al.} \cite{Engels2007} where the radial confinement  of an elongated
trap is periodically modulated in time.  By using a hydrodynamic description
of the superfluid Fermi gas, we first treat the case of a infinite
cylindrical gas to obtain an analytical expression for the spatial
modulation of the Faraday pattern as a function of the driving
frequency $\Omega$, for small frequencies $\Omega\ll\omega_\perp$,
$\omega_\perp$ being the transverse trap frequency.
Then we numerically solve the three-dimensional hydrodynamic equations
taking into account the axial confinement and investigate the
formation of Faraday waves at higher frequencies ($\Omega\sim 2
\omega_\perp$) where the fringes in the density profile can be
experimentally visualized in cigar-shaped traps.  Finally we compare
the analytical prediction for the infinite cylinder with the numerical
results and find that our analytical formula provides a good estimate
of the pattern modulation even at $\Omega\sim 2\omega_\perp$.

The manuscript is organized as follows. Section \ref{sec:model}
introduces the hydrodynamic description of the gas and derives the low
frequency solution of the spatial pattern formation. In Sec. \
\ref{sec:num} we present the numerical results for the dynamics in a
cigar-shaped trap and compare them with our analytical prediction.
Section \ref{sec:summ} offers a summary and concluding remarks.

\section{\label{sec:model}Faraday excitations in a cylinder: 
hydrodynamic description}
Let us consider a superfluid Fermi gas at $T=0$ confined by a cylindrical 
potential $V_{\perp}(\mathbf{r})=m\omega_\perp^2r^2/2$ with
$r^2=(x^2+y^2)$ and $m$ the atom mass.  The dynamics of the density
profile $n(\mathbf{r},t)$ and   the velocity field ${\bf v}(\mathbf{r},t)$ is
described by superfluid hydrodynamic equations, namely, the continuity equation
\begin{equation}
\partial_t n+{\bf\nabla}\cdot(n{\bf v})=0
\label{continuity}
\end{equation} 
and the Euler equation
\begin{equation}
m\partial_t {\bf v}+{\bf\nabla}\left[\mu(n)+\dfrac{1}{2}m\omega_\perp^2r^2+\dfrac{1}{2}m{\bf v}^2\right]=0,
\label{Euler}
\end{equation}
where $\mu(n)$ is the equation of state (EOS) of the gas which
contains the microscopic details of the superfluid.

A solution of the dynamics for a time-dependent
$\omega_{\perp}$ can be obtained by relying on a scaling Ansatz.
Following Ref.~\cite{Kagan1996} we introduce the scaling
parameter $b(t)$ and write
\begin{equation}
n(r,z,t)=\dfrac{1}{b(t)^2}n_0(r/b(t))\qquad \text{and}\qquad
{\bf v}=\dfrac{\dot{b}(t)}{b(t)}{\bf r},
\label{scaling}
\end{equation}
where $n_0(r)$ is the stationary solution of the Euler equation at
$t=0$, i.e., the Thomas-Fermi (TF) profile defined by
$\mu(n_0(r))=\left[ \bar\mu -V_{\perp}(r)\right]$ with
$\bar\mu$ the chemical potential. 
The scaling Ansatz in Eq. (\ref{scaling}) satisfies the continuity
equation for any form of $b(t)$ and is an exact solution of the Euler
equation if the EOS follows a power law, $\mu(n)=\mathcal{C}n^\gamma$. In
this case the scaling parameter obeys the differential equation
\begin{equation}
\ddot{b}+\omega_\perp^2(t)\,b-\dfrac{\omega_\perp^2}{b^{2\gamma+1}}=0.
\label{scal}
\end{equation} 
Notice that if the EOS is not a power law, one can still introduce an
effective exponent $\bar\gamma$ and exploit the same procedure
\cite{Hui2004} to find an approximate solution of the dynamics.

To excite Faraday waves, we modulate the transverse trap frequency as
\begin{equation}
\omega_{\perp}(t) = \omega_{\perp}(1+\varepsilon\sin\Omega t)
\end{equation}
where $\Omega$ is the driving frequency and $\varepsilon$ is a small
  parameter, $\varepsilon \ll 1.$
For $\Omega\ll\omega_\perp$, Eq. (\ref{scal}) has the 
solution \cite{Kagan2007a}
\begin{equation}
b(t)\simeq 1-\alpha\sin(\Omega t)
\label{eq:bsim}
\end{equation}
with $\alpha=\varepsilon/(\gamma+1 -\Omega/(2\omega_{\perp}) )\simeq
\varepsilon/(\gamma+1)$
in agreement with the expression found in Ref. \cite{Kagan2007} for an
ultracold Bose gas ($\gamma=1$). For larger $\Omega$, the
solution of Eq.  (\ref{scal}) represents a forced breathing mode which
does not adiabatically follow the forcing frequency. For instance,
when the driving frequency is close to the  natural breathing mode
$\Omega=\sqrt{2\gamma+2}\,\omega_{\perp}$, the system exhibits the
resonant solution \begin{equation} b(t) \simeq 1 +
\frac{\varepsilon\,\omega_{\perp}t}{\sqrt{2+2\gamma}}\cos(\Omega t)
\label{eq:bsimres} \end{equation} valid for small $t$. For larger $t$,
the resonant solution can be seen to exponentially increase. 

\subsection{Excited states}
The modulation of the nonlinearity excites sound-like phonons in the
$z$-direction which, in turn, lead to the instability of the scaling
solution (\ref{scaling}).  These excited states breaking the axial
symmetry can be investigated by a linear analysis, i.e., by setting 
$n(r,z,t)=n_0(r/b)/b^2+\delta
n(r,z,t)$ and  linearizing Eqs. (\ref{continuity}) and
(\ref{Euler}) in $\delta n$ we arrive the following equation
\begin{equation}
m\partial^2_t\delta
n=\dfrac{1}{b^{2\gamma}}{\bf\nabla}\cdot\left[n_0\left(\frac{r}{b}\right)
{\bf\nabla}\left(\left.\dfrac{\partial\mu}{\partial
n}\right|_{n=n_0\left(\frac{r}{b}\right)}\!\!\hspace*{-1em}\delta n\right)\right],
\label{eqhydro}\end{equation}
that is valid for small velocity fields, i.e., 
$\Omega\ll\omega_\perp$. Equation (\ref{eqhydro}) is 
a sound-wave equation for the instantaneous density.
Using the solution Eq.\ (\ref{eq:bsim}) and developing to first order in
$\alpha$, Eq. (\ref{eqhydro}) gives rise to 
\begin{eqnarray}
m\partial^2_t\delta n&=&{\bf\nabla}\cdot\left[n_0|_{\alpha=0}
{\bf\nabla}\left(\left.\dfrac{\partial\mu}{\partial n}\right|_{\alpha=0}\!\!\delta n\right)\right]\nonumber\\
&+&2\alpha\sin (\Omega t)\,{\mathcal{F}}(n_0
|_{\alpha=0},\mu(n)_{\alpha=0},\delta n)
\label{espando}
\end{eqnarray}
with
\begin{eqnarray}
\mathcal{F}&=&{\bf\nabla}\cdot\left[\gamma n_0
{\bf\nabla}\left(\dfrac{\partial\mu}{\partial n}\delta n\right) -
\dfrac{r^2}{R_\perp^2}\dfrac{\partial n}{\partial \mu}{\bf\nabla}\left(\dfrac{\partial\mu}{\partial n}\,\delta n\right)\right.\nonumber\\
&-&\left.n_0{\bf\nabla}\left(\dfrac{r^2}{R_\perp^2}\,\dfrac{\partial n}{\partial\mu}\,\dfrac{\partial^2\mu}{\partial n^2}\,\delta n
\right)\right]_{\alpha=0}.
\end{eqnarray}
where $R_{\perp}=\sqrt{2\bar{\mu}/m\omega_{\perp}^2}$ is the
TF radius of the gas.  The right-most term in Eq.
(\ref{espando}) proportional to $\sin(\Omega t)$ behaves like a source
for excitations in the linear response regime.  Expressing $\delta n$
in its Fourier components, $\delta n=\sum_j\delta n_j(r,z)
\cos(\omega_jt+\phi_j)$, one can easily demonstrate that the
lowest-frequency Fourier component that is resonant with the
perturbation corresponds to $\omega_0=\Omega/2$. This  is a
general result for dynamical systems governed by Mathieu-type
equations \cite{rev} as Eq.\ (\ref{espando}).  We keep only the resonant term
in Eq.~(\ref{espando}) and make the Ansatz $\delta n_0(r,z)=\delta
n_0(r)\cos(qz)$, corresponding to counter propagating axial phonons
with momenta $\pm q$.  To find an estimate of $q$, we assume that
$\Omega$ is small enough to consider only the sound mode $\delta n_0$
and therefore take $ \delta n_0(r)=\left({\partial\mu}/{\partial
n}\right)^{-1}$ evaluated at $\alpha=0$, i.e.,  the sound limit
solution for $q\rightarrow 0$ and $\Omega\rightarrow 0$ at $t=0$ (see
Ref. \cite{Capuzzi2006a}).  By integrating in the radial plane one
obtains
\begin{equation}
m\left(\dfrac{\Omega}{2}\right)^2\int\left(\left.\dfrac{\partial\mu}{\partial
n}\right|_{\alpha=0}\right)^{-1}{\rm d}^2r=
q^2\int n_0|_{\alpha=0}{\rm d}^2r+O(\alpha^2)q^2
\label{eq1}
\end{equation}
which, for a power-law EOS leads to
\begin{equation}
q=\dfrac{\Omega}{2}\sqrt{\dfrac{m(\gamma+1)}{\gamma\bar\mu}},
\label{eq2}
\end{equation}
neglecting second-order terms in $\alpha$. The above expression can be
directly linked to the sound velocity $c_s$ \cite{Capuzzi2006a} in an elongated superfluid
by $q=\Omega/(2c_s)$. 

By introducing the wavelength of the Faraday pattern $d=2\pi/q$ it is
possible to evaluate the number of fringes $\mathcal{N}$ in an
elongated superfluid confined by an harmonic potential
$V_{\text{ext}}(r,z) = V_{\perp}(r) + m\omega_z^2 z^2/2$.  Given that the
axial extent of the density can be written as $L_z=2R_\perp/\lambda$
where $\lambda=\omega_z/\omega_{\perp}$ and neglecting  edge effects,
Eq.\ (\ref{eq2}) yields 
\begin{equation}
\mathcal{N}=\dfrac{L_z}{d}=\dfrac{1}{\pi\lambda}\sqrt{\dfrac{\gamma+1}{2\gamma}}
\dfrac{\Omega}{\omega_\perp}.
\label{frange}
\end{equation}
For $\gamma=1$, $\mathcal{N}$ is in good agreement with the 
experimental data of
Engels and coworkers \cite{Engels2007} even for large
$\varepsilon$ and $\Omega$, where the procedure 
we used to derive it
is not fully justified.  Equation (\ref{frange})
predicts that the number of Faraday fringes varies by a factor of
$2/\sqrt{5}$  in the BCS-BEC crossover and thus it could be easily
observed in current experimental setups with strongly elongated traps
\cite{Joseph2007,Tarruel2007}, as discussed in the following section.
%The overall behaviour of the ratio $L_z/d$ as a function of $1/k_Fa$, $k_F$
%being the Fermi wave-number and $a$ the $s$-wave scattering length, for the case $\Omega=...$, is depicted in Fig. \ref{fig1}.
\section{\label{sec:num}Numerical results}
The numerical solution of the superfluid hydrodynamic equations allows to
fully take into account the role of the axial confinement and to get
over the approximation of small frequencies.
To compute the dynamics we first map the superfluid density
$n(\mathbf{r},t)$ and velocity field $\mathbf{v}(\mathbf{r},t)$ to the 
complex field $\psi(\mathbf{r},t)= \sqrt{n}\,e^{iS}$, with
$\mathbf{v}=\hbar\nabla S/m$. In this way the dynamical
equations can be cast in a Gross-Pitaevskii-like equation of motion
for $\psi$. We then discretize this equation in the $(r,z)$ plane and solve
it using a 5th-order Runge-Kutta method \cite{nr}. A mesh of
$200\times400$ points in the $(r,z)$ plane and a time step
$dt=10^{-4}/\omega_{\perp}$ are usually enough to compute the
evolution until $t=500/\omega_{\perp}$.

We consider a Fermi gas with $N=5\times10^5$ $^{40}$K atoms trapped in
a cylindrically symmetric trap with $\omega_{\perp}=160.5\,$Hz and
$\omega_z=7\,$Hz, and fix the modulation parameter $\varepsilon=0.05$.
The superfluid is described by the EOS developed by Hu \textit{et
al.} \cite{Hu2005} and parametrized in terms of $1/(k_F\,a)$ where $k_F$
is the Fermi wavevector $k_F=(3\pi^2\,n)^{1/3}$ and $a$ is the
$s$-wave scattering length.  If not differently specified, in the
following we refer to a continuous modulation of the transverse trap
frequency at $\Omega=2\omega_{\perp}$.

\begin{figure}
\includegraphics[width=0.95\columnwidth,clip=true]{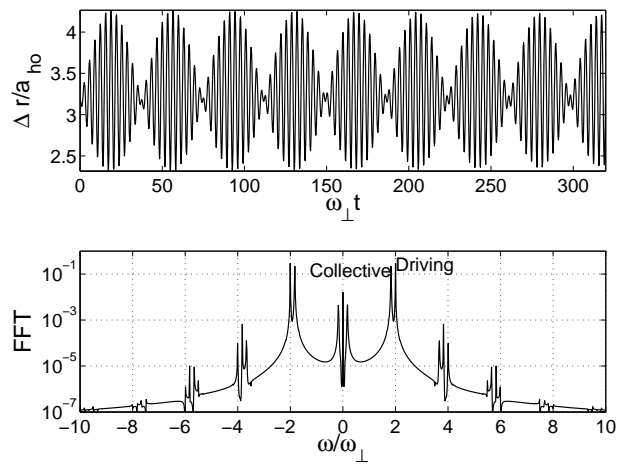}
\caption{\label{fig:dr}Time evolution of the transverse width $\Delta
r$ for a Fermi superfluid in the BCS limit (see text). The top panel
depicts $\Delta r$ (in units of
$a_{\text{ho}}=\sqrt{\hbar/(m\omega_{\perp})}$) as a function of $t$
(in units of $\omega_{\perp}^{-1}$) and the bottom panel depicts its
Fourier transform (in arbitrary units and logscale) as function of
$\omega/\omega_{\perp}$.} 
\end{figure}

\begin{figure}
\includegraphics[width=0.95\columnwidth,clip=true]{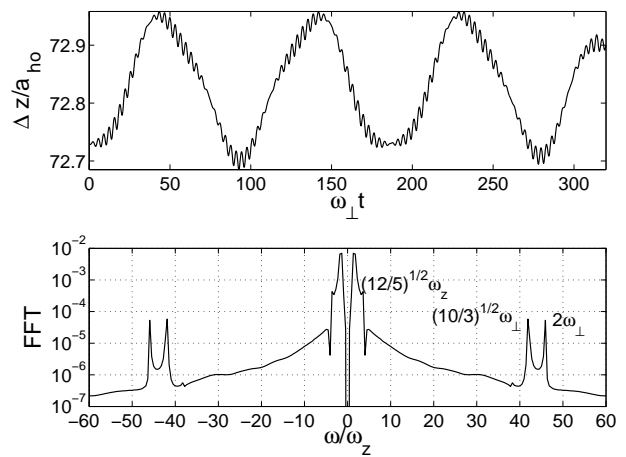}
\caption{\label{fig:dz}Idem Fig.\ \ref{fig:dr} for the axial
width $\Delta z$. The Fourier transform in the bottom panel is
displayed in arbitrary units and logscale, and the frequency $\omega$
in units of $\omega_z$.}
\end{figure}

\begin{figure}
\includegraphics[width=0.9\columnwidth,clip=true]{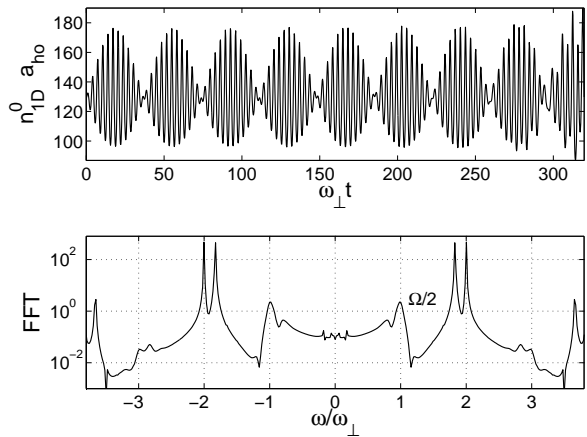}
\caption{\label{fig:rho0t} Time evolution of the column density at the
trap center $n_{\text{1D}}^0=n_{\text{1D}}(z=0)$. The top panel shows
$n_{\text{1D}}^0$ (in units of $a_{\text{ho}}^{-1}$) as a function
of time $t$ (in units of $\omega_{\perp}$). The bottom panel
displays its temporal Fourier transform (in arbitrary units and
logscale) as function of $\omega/\omega_{\perp}$.}
\end{figure}

From the numerical solution of the density profile $n(r,z,t)$ we have
computed the axial and radial RMS widths, and the column densities
$n_{\text{1D}}(z,t)=\int \text{d}y\,n(x=0,y,z,t)$.   These are shown
in Figs.  \ref{fig:dr}-\ref{fig:rho0t} together with their Fourier
transforms for the case of the Fermi gas in the BCS regime
($a/a_0=-10^{-1}$, $a_0$ being the Bohr radius). The dynamics of the width
$\Delta r$ (Fig.\ \ref{fig:dr}) is dictated by the driving frequency
coupled to the transverse breathing mode, which for the gas in the BCS
limit is $\omega=\sqrt{10/3}\, \omega_{\perp}$.  This reflects the
competition between the forcing and the natural breathing mode.
Furthermore, within the scaling approach of Sec.\ \ref{sec:model}, the
axial width evolves in time with $b(t)$, i.e., $\Delta r(t)=\Delta
r(0)\,b(t)$.  Indeed, the numerical solution of the ordinary
differential equation (\ref{scal})  reproduces very well the behavior
of $\Delta r$ for the axially confined gas before the Faraday
instability sets in. As noted earlier, we would like to emphasize that
due to the vicinity of the natural breathing mode and the driving
frequency, $\Delta r$ does not simply adiabatically follows
$\omega_{\perp}$ as clearly shown in the figure.

The time evolution of the axial width $\Delta z$ and
its Fourier transform are instead illustrated in Fig. \ref{fig:dz}.
Similarly to the radial width, the evolution of $\Delta z$ is
dominated by the forcing frequency $\Omega$ and the axial breathing
mode, that is $\sqrt{12/5}\omega_\perp$ in the BCS case.
The excitation of the axial breathing mode is a consequence of the
coupling of the radial and axial directions imposed by the
three-dimensional trap and should not be expected in the infinitely
long cylinder  discussed in Sec.\ \ref{sec:model}.

Although  $\Delta r$ and $\Delta z$ show no evidence of the frequency
$\Omega/2$ in their spectra, the column density at the center of the
trap (Fig.\ \ref{fig:rho0t}) clearly exhibits it. The relative
importance of this frequency in the spectrum  of $n_{\text{1D}}$
is accentuated with time as a result of the formation of the Faraday wave,
this effect can be seen by analyzing the Fourier transform for
different time intervals.

\subsection{Onset of Faraday wave formation}
\begin{figure}
\includegraphics[width=0.95\columnwidth,clip=true]{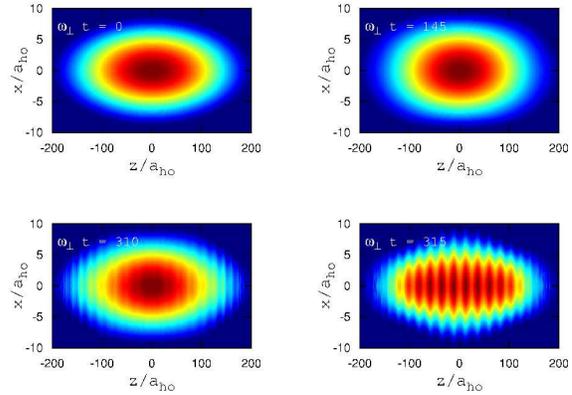}
\caption{\label{fig:rhotimeBCS} (color online) Density plots 
$n(x,y=0,z,t)$ of a Fermi gas in the BCS regime at several times $t$
during the transverse frequency modulation.} 
\end{figure}

The Faraday instability becomes visible when the momentum distribution
develops components at the momenta $\pm q$ corresponding to pairs of
counter-propagating axial excitations of frequency $\Omega/2$.  In
Fig. \ref{fig:rhotimeBCS} we display the formation of a Faraday wave
in the density profile of a gas in the weakly BCS limit with
$a=-10^{-1}a_0$. After approximately 50 driving periods, the
appearance of transverse fringes along the axis of the cigar is clear
and we can say that the Faraday instability has been fully set in. For
longer times, as the Faraday waves increase their amplitude, a large
number of  modes interacts and the Faraday pattern develops several
complicated spatial and temporal structures, resembling those of
spatio-temporal chaos as expected in confined hydrodynamic flows
\cite{Cross1993}. This regime cannot be accounted for by the
simplified analysis of Sec. \ref{sec:model} and requires a more
demanding numerical effort. 

%This regime, called spatio-temporal chaos, is intermediate between chaos 
%and turbulence. It is not chaos because many and not just a few degrees 
%of freedom cooperate, and it is not turbulence because it lacks the 
%large range of scales that is typical for strong turbulence.(reaching a cahotic regime? \cite{chaos}). CHECK!!!!!
%the dynamics of the system become more and
%more "chaotic"\cite{chaos}. (e' vero che sia caotica?)

\begin{figure}
\includegraphics[width=0.95\columnwidth,clip=true]{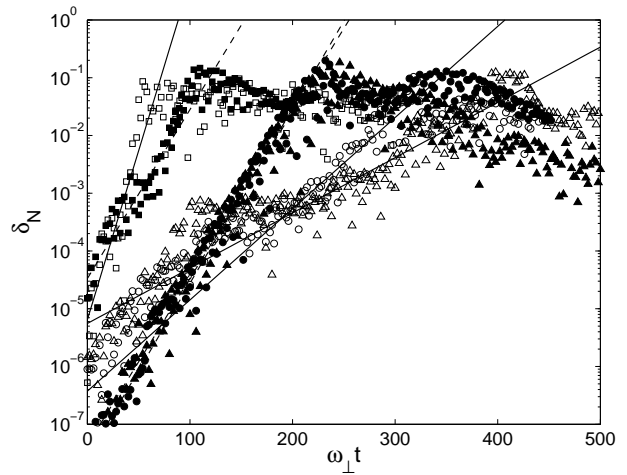}
\caption{\label{deltaN} Fraction $\delta_N$ (in logscale) 
as a function of the time $t$ (in units of $\omega_\perp^{-1}$) 
for  $a/a_{0}=-10^{-1}$ (circles) ,$-10^{5}$ (triangles), and $10^{3}$
(squares). The empty symbols corresponds to the continuous driving
while the full ones to the 3-cycle driving. The lines are guides to
eyes for better visualizing the growth rates.}
\end{figure}

The onset time of the Faraday waves varies through the BCS-BEC crossover.
%($\clubsuit$ oppure perche' e piu
%denso??? Direi di no. se fosse la densita' lo unitary dovrebbe essere
%sensibilmente piu' veloce del BCS$\clubsuit$).  
The growth rate of the spatial pattern can be evaluated by computing the
fraction 
\begin{equation}
\delta_N(t) = \left(\frac{\int \delta \tilde{n}^2(z,t)\, dz}{\int
\tilde{n}_0^2(z,t)\,dz}\right)^{1/2}
\end{equation}
where $\tilde{n}_0(z,t)$ is the scaled density extracted from the
numerical solution by fitting a TF profile $n(z) =
\mathcal{A}(\tilde{\mu}-m\omega_z^2z^2/2)^{\eta}$ to the integrated
density $n_{\text{1D}}(z,t)$ at each time $t$, and $\delta
\tilde{n}= n_{\text{1D}}(z,t)-\tilde{n}_0(z,t)$.  The magnitude
$\delta_N$ represents the normalized deviation of the density
respect to the scaling solution and can therefore be related to the
amplitude of the Faraday wave. The evolution
of $\delta_N$ is illustrated in Fig. \ref{deltaN} for three values
of $a$ in the BCS, unitarity and BEC regimes. 
In the BCS and in the unitarity limits the growth rates are comparable
as the excitation spectra in both regimes are the same. In the BEC
limit, instead, the onset is more rapid since the driving frequency
$\Omega=2\omega_{\perp}$ corresponds to a collective mode of the
cloud. 

Faraday waves can be also created by driving the transverse
confinement for only few cycles \cite{Engels2007}.  However, in this
case, the growth rate of the pattern is modified.  For instance, we
have found that for a 3-cycle perturbation, the rate for a
BCS and a unitarity superfluid Fermi gas increases approximately by
a factor of 2--3, while in the BEC limit it decreases by a factor of
1.5 approximately.
For our choice of $\Omega$, the driving frequency and the breathing
modes are in competition in the BCS and unitarity limits and the
continuous excitation may delay the formation of the pattern. In the BEC
limit, $\Omega=2\omega_\perp$ corresponds to the breathing mode and
the onset of the instability is more rapid by continuously exciting
the system at resonance.  Note that  the total growth time
of a Faraday wave may also depend on the numerical noise caused by 
the discrete space-time grids, but the growth rates in the linear
regime should not be modified.

\subsection{Faraday wavevector through the crossover}
\begin{figure}
\includegraphics[width=0.95\columnwidth,clip=true]{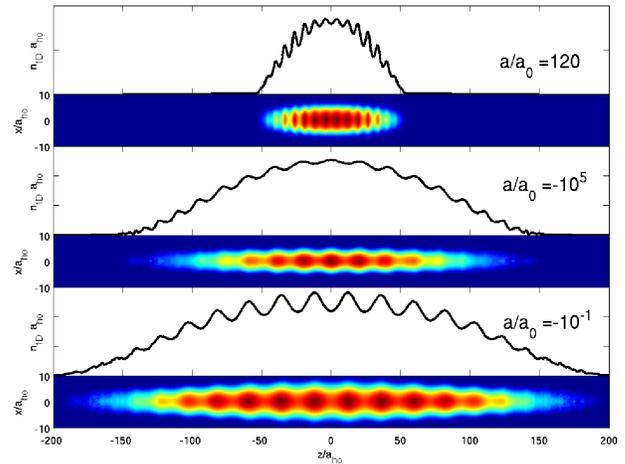}
\caption{\label{fig:rhozpanel}(color online) Density plots and integrated density
$n_{\text{1D}}$ 
as function of position  (in units of $a_{\text{ho}})$ after the
Faraday instability has been set in. Top, middle, and bottom panels
correspond to $a/a_0=120,-10^5$, and $-10^{-1}$, respectively.}
\end{figure}

A further observable which depends on the crossover is the wavevector
$q$ of the Faraday pattern, as already outlined in Sec.
\ref{sec:model}.  Experimentally, $q$ can be deduced by measuring the
interfringe distance $d$ at the center of the cloud. 
The interfringes  obtained in the numerical simulation can be observed
in Fig.\ \ref{fig:rhozpanel} where the integrated density
$n_{\text{1D}}$ has been plotted together with the topview density
plots for $a/a_0=-10^{-1},-10^5$, and $120$ corresponding to the BCS,
unitarity and BEC regimes, respectively.  It is worth stressing that
at $a=120 a_0$, the transverse size displays an impact-oscillator
behavior as observed for an atomic BEC at the breathing frequency
\cite{Engels2007,Nicolin2007}. The resulting strong modulation of the
nonlinearity speeds up formation of the Faraday waves
and impose severe numerical limitations.
%that is why for this case we
%used a spatial grid with 1600$\times$600 points in the $(r,z)$ plane
%an a $dt = 10^{-5}/\omega_{\perp}$.

Through the crossover, from the BCS to the BEC limits, at
fixed number of fermions, both the size of the cloud and the
interfringe $d$ decrease since they are both proportional to
$\sqrt{\bar\mu}$; however, $d$ decreases less rapidly
because of the factor $\sqrt{\gamma/(1+\gamma)}$ with a
corresponding slight modification of the fringe number. As a result
in the BCS and unitary regimes the number of
fringes are the same as can be seen in Fig.\ 6, i.e., 16 for $a=-0.1 a_0$ and
$-10^5 a_0$, while in the BEC regime ($a=120 a_0$), the number of fringes
diminishes by a factor of $2/\sqrt{5}$, corresponding to 14 fringes
approximately.  In more elongated traps as that used in the ENS
experiments \cite{Tarruel2007} with $\lambda\simeq 0.18$, the
variation of $\mathcal{N}$ through the crossover can be more
important, from $\mathcal{N}= 41$ in the BCS limit to $\mathcal{N}=37$
in the BEC limit at $\Omega=2\omega_\perp$.

\begin{figure}
\includegraphics[width=0.95\columnwidth,clip=true]{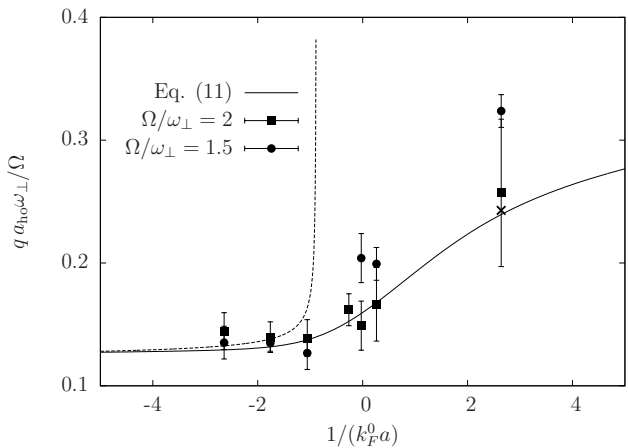}
\caption{\label{fig:sound_comparison} Comparison of the ratio
$q/\Omega$ between the observed Faraday wavevector $q$ (in units of
$a_{\text{ho}}$) and the driving frequency $\Omega$ (in units of
$\omega_{\perp}$) with that predicted for a soundlike excitation (Eq.
(\ref{eq1})) as functions of $1/(k_F^0a)$, where $k_F^0$ is the Fermi
momentum of the noninteracting gas at the trap center.  The error
bars have been estimated from  the variance of $d$ during several time
steps. The cross corresponds to a result obtained by modulating at
$\Omega/\omega_{\perp}=1.5$ corrected by taking into account the
induced excitation at $\Omega=2\omega_{\perp}$ (see text). The dashed
line corresponds to the prediction of Eq.(\ref{eq1}) with the BCS
mean-field EOS.} 
\end{figure}

In Fig. \ref{fig:sound_comparison} we compare the Faraday wavevector
$q=2\pi/d$ as observed in the numerical simulation with that predicted
for a soundlike excitation at $\Omega/\omega_{\perp}=2$ and $1.5$.
Mimicking a real experiment, $d$ has been measured at the center of
the density profile and the error bars have been evaluated from  its
variance during several time steps.  Equation (\ref{eq1}) instead has
been evaluated by using the EOS deduced by Hu and coworkers
\cite{Hu2005}. The results show that Eq. (\ref{eq1}) 
provides a good estimate of the function $q(\Omega)$ through all the
crossover in particular for $\Omega=2\omega_{\perp}$. On the other
hand, for $\Omega=1.5\omega_{\perp}$ the numerical result for
$a=10^3\,a_0$ lies well above the prediction. Indeed, by analyzing 
the FFT of $n_{\text{1D}}^0$ in this case, we found that the transverse
modulation at $\Omega=1.5\omega_{\perp}$  also excites  the 
collective mode of the BEC at $\Omega=2\omega_{\perp}$ similarly to
what was found experimentally in Ref.\ \cite{Engels2007}. As a
consequence,  the mode at $\Omega=2\omega_{\perp}$ dominates the
generation of longitudinal excitations even when modulating at
$\Omega=1.5\omega_{\perp}$ and thus the numerical result should be
compared with the prediction at $\Omega=2\omega_{\perp}$, as shown by
a cross in Fig.\ \ref{fig:sound_comparison}. For comparison, we have
also evaluated the Faraday wavevector predicted by  Eq. (11) using the
mean-field BCS EOS,
$\mu(n)=\hbar^2k_F^2/2m\,(1+\tfrac{4}{3\pi}\,k_Fa)$. It is worthwhile
remembering that this EOS predicts the collapse of the fermionic
mixture for strong attractive interactions, as indicated by the
divergence of $q$ shown in dashed line in Fig.\
\ref{fig:sound_comparison}.

\section{\label{sec:summ}Summary and concluding remarks}
In summary we have studied the formation of Faraday patterns in an
elongated superfluid Fermi gas in the BCS-BEC crossover.  We simulated
an experiment where nonlinearity is driven by modulating the
transverse confinement and compared the numerical results with an
analytical estimate derived neglecting the axial confinement
and for low frequencies.  The peculiarity of a superfluid
Fermi gas respect to an atomic Bose-Einstein condensate is that the
nonlinear term controls the microscopic features of the superfluidity
and that the pattern modulation depends on the crossover itself.
Moreover the pattern growth rate varies through the crossover as a
consequence of the different spectra.  Our numerical and analytical
predictions may be observed in current experiments with superfluid
Fermi gas in strongly elongated traps \cite{Joseph2007,Tarruel2007}.

\begin{acknowledgments}
This work was partially supported by grants PICT 31980/05 from ANPCyT and PIP
5138/05 from CONICET, Argentina. P.C. wishes to acknowledge the kind
hospitality of the INLN where most of this work was performed.
\end{acknowledgments}

\end{document}